\documentclass[conference]{IEEEtran}
\usepackage{amsmath,amsfonts}
\usepackage{amssymb}
\usepackage{algorithmic}
\usepackage{array}
\usepackage{textcomp}
\usepackage{stfloats}
\usepackage{url}
\usepackage{verbatim}
\usepackage{graphicx}
\usepackage{cite}
\usepackage{subfigure}
\usepackage[linesnumbered,ruled,vlined]{algorithm2e} 
\usepackage{color}
\usepackage{booktabs} 
\usepackage[table,xcdraw]{xcolor}
\usepackage{multirow}
\usepackage{enumitem}
\usepackage{makecell}
\usepackage{bm}
\usepackage[colorlinks,linkcolor=blue,citecolor=blue]{hyperref}
\begin{document}
\setlength{\textfloatsep}{2pt} 
\setlength{\skip\footins}{5pt}

\title{Accelerating Multi-Condition T2I Generation via Adaptive Condition Offloading and Pruning}

\IEEEoverridecommandlockouts
\author{
	\IEEEauthorblockN{
		Yuxin Kong\IEEEauthorrefmark{2}, 
		Peng Yang\IEEEauthorrefmark{2}\IEEEauthorrefmark{1}, 
		Chongbin Yi\IEEEauthorrefmark{2}, 
		Fan Wu\IEEEauthorrefmark{3}, 
		and Feng Lyu\IEEEauthorrefmark{3}
    } 
	\IEEEauthorblockA{\IEEEauthorrefmark{2} School of Electronic Information and Communications, Huazhong University of Science and Technology, Wuhan, China}
	\IEEEauthorblockA{\IEEEauthorrefmark{3}School of Computer Science and Engineering, Central South University, Changsha, China}
  Email: \IEEEauthorrefmark{2}\{yxkong, yangpeng, chongbin\_y\}@hust.edu.cn. \IEEEauthorrefmark{3}\{wfwufan, fenglyu\}@csu.edu.cn.
  \thanks{\IEEEauthorrefmark{1}Peng Yang is the corresponding author.}
} 

\maketitle
\begin{abstract}

Text-to-image (T2I) generation using multiple conditions enables fine-grained user control on the generated image. Yet, incorporating multi-condition inputs incurs substantial computation and communication overhead, due to additional preprocessing subtasks and control optimizations. It hence leads to unacceptable generation latency.
In this paper, we propose an end-edge collaborative system design to accelerate multi-condition T2I generation through adaptive condition offloading and pruning.
Extensive offline profiling reveal that, different conditions exhibit significant diversity in computation and communication costs. To this end, we propose a \textit{Subtask Manager} that jointly optimizes condition inference offloading and bandwidth allocation using a heuristic algorithm, balancing local and edge execution delays to minimize overall preprocessing latency.
Then, we design a lightweight feature-driven \textit{Conditioning Scale Estimator} that evaluates the contribution of each condition by analyzing its feature activation strength and overlap with other conditions. This allows adaptive conditioning scale selection and pruning of insignificant conditions, thereby accelerating the denoising process.
Extensive experimental results show that our system reduces latency by nearly 25\% and improves 6\% average generation quality, outperforming other benchmarks.

\end{abstract}








\section{Introduction}\label{secintro}

\begin{figure}[t]
   \centering
   \includegraphics[scale=0.46]{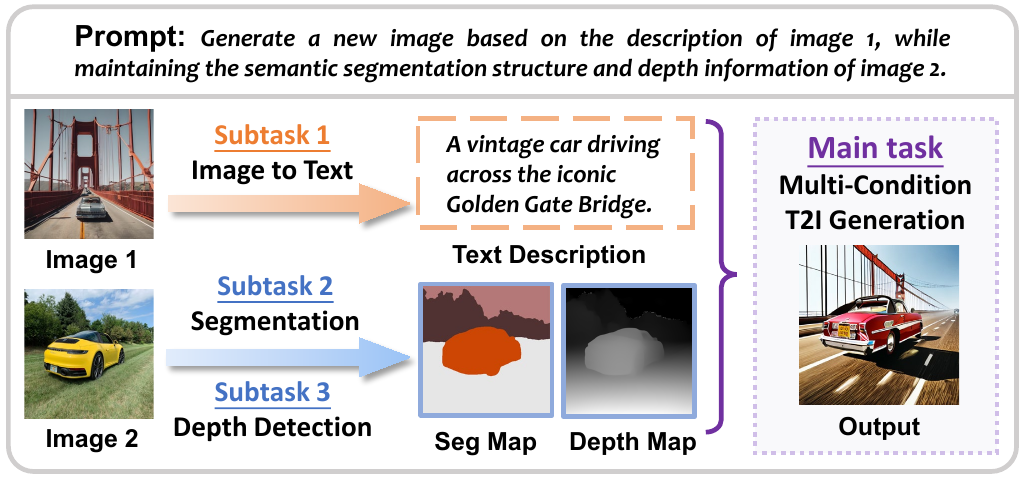}
   \caption{An example of multi-condition T2I generation task.}
   \label{figintro}
   
\end{figure}

Driven by the rapid advancement of diffusion models \cite{stadiff}, text-to-image (T2I) generation has emerged as a prominent application of Artificial Intelligence Generated Content (AIGC). It allows users to easily create visually stunning images simply by entering a text prompt \cite{dreambooth}. Yet, as users increasingly demand finer-grained control over generated images, embedding text prompts alone become insufficient. This demand gives rise to \textit{multi-condition T2I generation} \cite{cocktail}, a paradigm that jointly utilizes multiple auxiliary control conditions extracted from user-provided reference images to guide the generation process \cite{comdos}. As illustrated in Fig. \ref{figintro}, a typical multi-condition T2I task begins with a compositional user prompt that comprises multiple conditions, which boil down to  several heterogeneous preprocessing subtasks, followed by a main generation task that integrates their outputs. In this way, the main generation model can only be executed after three prerequisite subtasks complete: \textit{(1)} extracting a textual description from Image 1, \textit{(2)} producing a semantic segmentation map from Image 2, and \textit{(3)} generating a depth map from Image 2. Upon completion of all the subtasks, the resulting control conditions are fed into the main model together, yielding a highly controlled and cohesive output. In contrast to conventional text-only image generation, this multi-condition paradigm greatly enhances compositional flexibility and better satisfies user's control intent.


However, T2I generation is too computationally demanding for efficient local execution on user devices, and integrating multiple control conditions further increases this burden. With the proliferation of resource-sufficient edge nodes in proximity to users, edge computing has become a promising paradigm for delivering diverse kinds of AIGC services. Existing studies have shown that this paradigm achieves remarkable efficiency for conventional text-only image generation by offloading the entire generation task to nearby edge servers, thereby avoiding the excessive costs of cloud-based solutions \cite{swtmc, icme25}. Yet, in multi-condition T2I generation, where multiple diversified conditions should be processed, the existing edge service paradigm faces the following three key challenges.

\textit{Firstly,} beyond the heavy denoising process of main task, each multi-condition T2I generation require extensive preprocessing subtasks to extract diverse conditions, substantially increasing the computational burden on the edge server. This issue is further exacerbated in multi-user scenarios, where concurrent generation requests from multiple users significantly increase the total preprocessing workloads, rapidly draining edge resources and extending service delays. While off-the-shelf user devices lack the computational power to perform the resource-intensive denoising process of the main task, they are generally capable of efficiently executing preprocessing subtasks locally \cite{yxtmc}. An adaptive subtask offloading strategy is therefore crucial to alleviate the computational load on the edge server and thereby improving system efficiency.






\textit{Secondly,} in addition to the extra preprocessing subtasks, multi-condition T2I generation also necessitates the activation of additional control modules during the denoising process. Each module entails considerable computation to inject the control conditions into the U-Net backbone through trainable layers for controlled generation \cite{controlnet}. Specifically, incorporating each additional ControlNet module for Stable Diffusion v1.5 enlarges the model size by approximately 42\% \cite{subject200k}, resulting in an extra inference latency of about 0.45 s per condition ($\thickapprox $38\% increase) with 30 denoising steps on an RTX 5080 GPU. As a result, the concurrent engagement of multiple control modules introduces considerable additional computation, significantly increasing inference latency. Thus, mitigating this burden to enable faster generation remains challenging.





\textit{Thirdly,} while text-only image generation suffices with uploading text prompts at negligible bandwidth cost, multi-condition paradigm necessitates transmitting larger preprocessed visual conditions or source images, resulting in higher communication overhead. Specifically, local preprocessing requires uploading the compact conditioning output, with data sizes varying across subtasks. In contrast, edge-only preprocessing involves transmitting large source images, leading to higher data volume that scales with multiple reference images required for different preprocessing subtasks. In multi-user scenarios, concurrent requests further exacerbate this issue by severely limiting available uplink bandwidth per user, which is typically much scarcer than the downlink bandwidth \cite{yxtccn}. As different subtask offloading decisions lead to different transmission demands, bandwidth resources should be carefully allocated among users to minimize overall transmission delays.

To address the above challenges, in this paper, we propose a resource-efficient system design for multi-condition T2I generation, enabling both low-latency and high-quality generation through end-edge collaboration. This system starts by collecting multi-condition T2I generation requests from users, each of which is parsed into multiple preprocessing subtasks and a main image generation task. Then, we propose a \textit{Subtask Manager} to jointly optimize subtask offloading decisions and bandwidth allocation across users via a heuristic algorithm balancing local-edge execution delays. Depending on the decisions, user-side and edge-side subtasks are executed in parallel. For offloaded subtasks, source images are uploaded to the edge for preprocessing, and the resulting control conditions are sent directly to the main task. For local subtasks, inference runs on the user device, after which the generated conditions are transmitted to the edge. Once all subtasks complete, we present a feature-driven \textit{Conditioning Scale Estimator}, which evaluates the contribution of each visual condition to determine optimal conditioning scales while pruning insignificant ones. Finally, the remaining conditions along with their estimated scales are fed into the main model for generation, achieving lower inference latency while preserving high generation quality. Our main contributions are summarized as follows.



\begin{itemize}
   \item We propose an end-edge collaborative system design for accelerating multi-condition T2I generation, featuring an algorithm that adaptively optimizes preprocessing subtask offloading and bandwidth allocation decisions to minimize subtask completion latency.
   \item We design a lightweight feature-driven method to estimate appropriate conditioning scales for visual conditions, while pruning insignificant ones to accelerate inference with superior generation quality.
   \item We validate our proposed system through extensive experiments, demonstrating 25\% reductions in overall latency and 6\% improvements in generation quality.
\end{itemize}


\begin{figure}[t]
  \centering
  \includegraphics[scale=0.48]{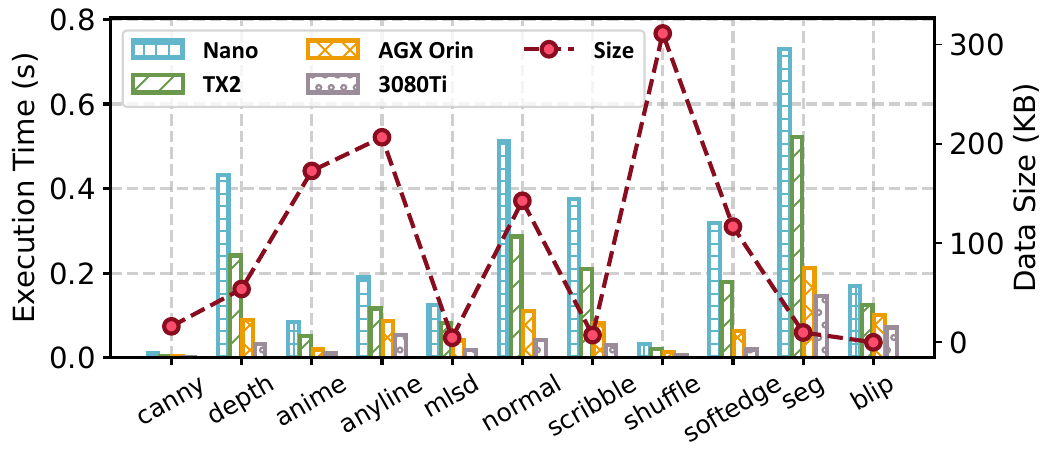}
  \caption{Heterogeneous computing and transmission demands across subtasks.}
  \label{figmoti1}
\end{figure}







\section{Motivation}\label{secmoti}


\subsection{Diversified Resource Demands of Subtasks}\label{secmoti1}

In multi-condition T2I generation, users must first preprocess the input images to obtain the corresponding control conditions, including visual conditions and textual prompts. Different control conditions rely on distinct preprocessing models, which exhibit diversified computing resource demands. Within the end-edge collaborative framework, these preprocessors can be executed either on the end device or offloaded to the edge. As a result, different offloading decisions incur different computation and communication overheads.

To investigate this diversity in resource demands, we evaluate 11 widely used preprocessors on representative end and edge platforms, including Jetson Nano, Jetson TX2, Jetson AGX Orin, and RTX 3080Ti GPU, by measuring their processing latency and output data size for 512*512 input images. 

The results, summarized in Fig. \ref{figmoti1}, reveal substantial disparities in the execution latency across different models, which are also strongly dependent on the underlying hardware. For instance, on the resource-constrained Jetson Nano, extracting semantic segmentation maps incurs significantly higher latency than extracting Canny edge maps, while the latter still requires approximately four times the processing time compared to execution on more powerful Jetson AGX Orin. Moreover, the results on the data size of control conditions indicate that, in most cases, the generated conditioning data are substantially smaller than the original input images, while still exhibiting considerable variation across different preprocessors.

When multiple conditions are jointly employed, their diversified resource demands make subtask offloading and bandwidth allocation both critical to overall latency, thereby necessitating a joint and adaptive optimization strategy.

\begin{figure}[t]
  \centering
  \includegraphics[scale=0.59]{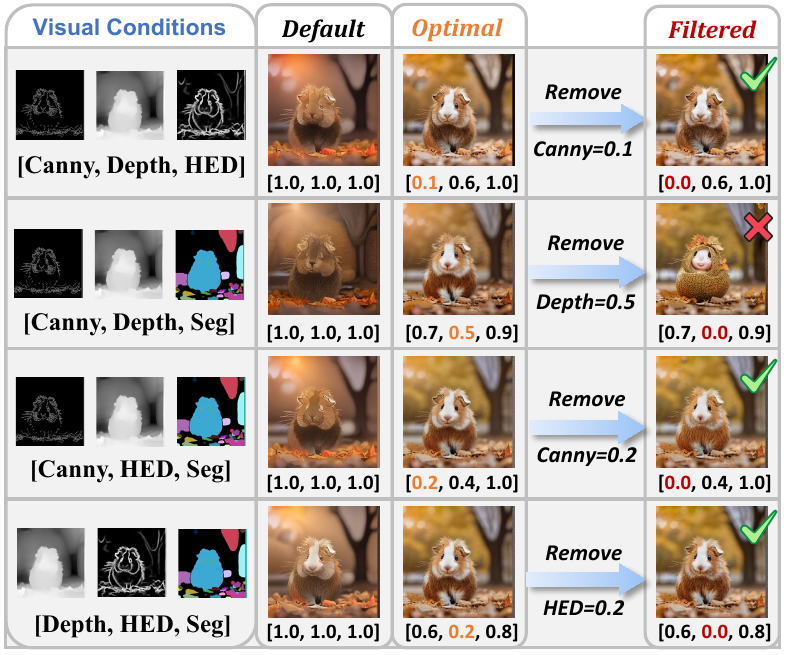}
  \caption{Impact of conditioning scales under different visual condition combinations with the prompt of \textit{It is sitting on a bed of autumn leaves with a blurred tree in the background, displaying a curious expression.}}
  \label{figmoti2}
\end{figure}

\subsection{Impact of Conditioning Scales}\label{secmoti2}

In multi-condition T2I generation, each visual condition is associated with a \textit{conditioning scale} that controls its influence during the denoising process by modulating feature injection from the corresponding ControlNet branch into the generative backbone \cite{controlnet}. Unless specified otherwise, all conditioning scales are set to 1.0 by default in standard Implementation.


We investigate the impact of conditioning scale configurations on image generation across different combinations of visual conditions. As shown in Fig. \ref{figmoti2}, we consider four types of visual conditions and jointly employ three for each generation. We then compare images generated under the default setting with those produced using the optimal configuration, defined as the combination of conditioning scales that maximizes the ImageReward\cite{imagereward} and DreamSim\cite{dreamsim} scores. The results show that default settings often yield suboptimal generation results, and achieving optimal quality requires fine-grained tuning of conditioning scale combinations. 

We further ablate the visual condition with the lowest scale by setting it to zero and removing its control module. When the scale is sufficiently small, the removal has negligible impact on generation quality while reducing inference time by about 0.46 s ($\thickapprox$18\% decrease) per generation on an RTX 5080 GPU with 30 denoising steps. Conversely, eliminating a high-scale condition causes substantial quality degradation (\textit{e.g.,} the second row), highlighting its indispensable contribution.

Therefore, to further accelerate multi-condition T2I generation without compromising quality, it is crucial to accurately estimate the conditioning scales for each visual condition and prune the unnecessary or insignificant ones.


\section{System Design}\label{secsys}


\subsection{System Overview}\label{secsys1}

\begin{figure}[t]
  \centering
  \includegraphics[scale=0.36]{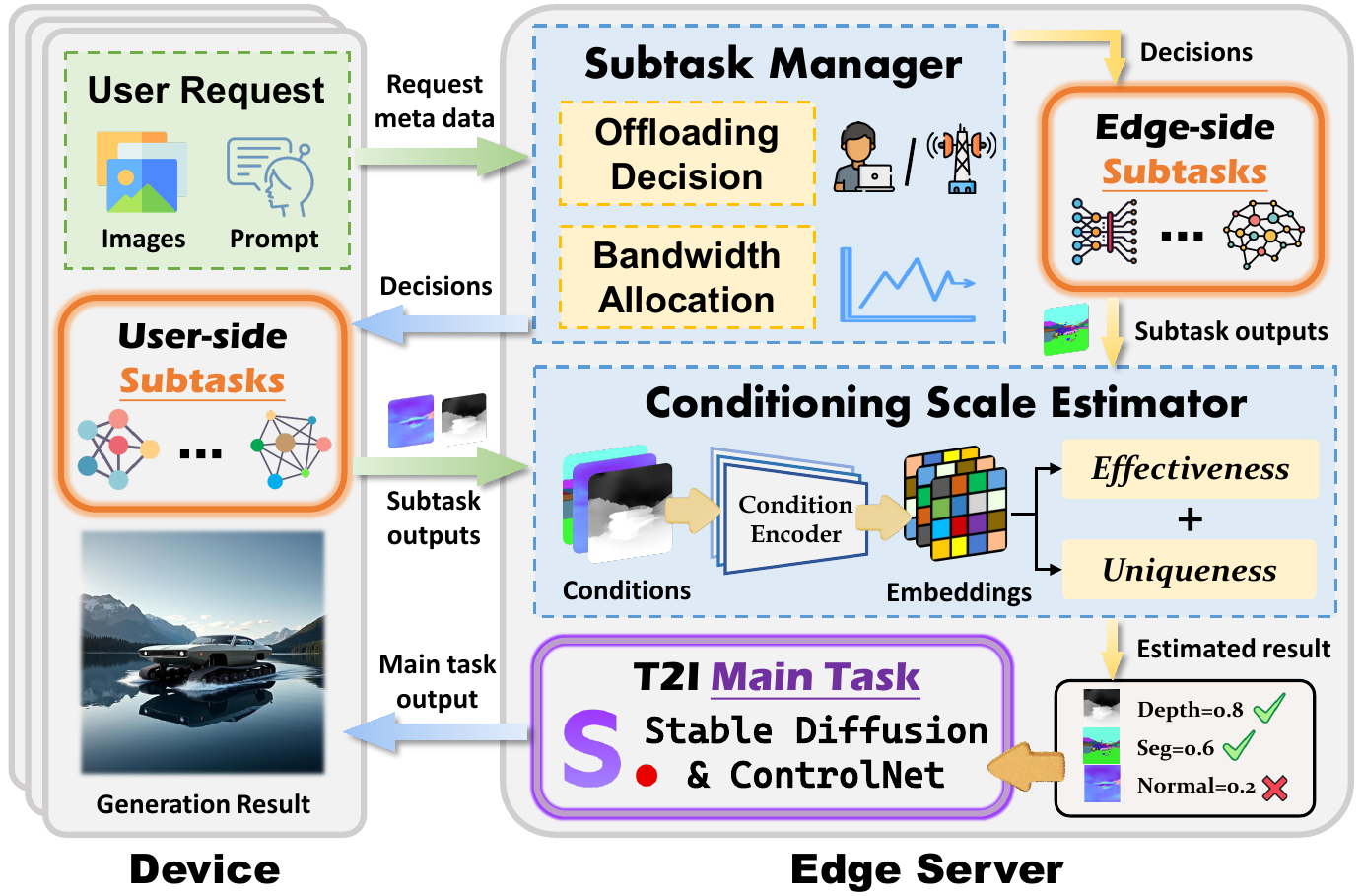}
  \caption{The proposed system design.}
  \label{figsys}
\end{figure}

As shown in Fig.~\ref{figsys}, we consider an end-edge collaboration scenario where a single edge server serves multiple users with diverse computing capabilities. Each user first submits a multi-condition T2I request prompt consisting of several preprocessing subtasks and a main image generation task to the edge. Then the \textit{Subtask Manager} jointly optimizes subtask offloading and multi-user bandwidth allocation decisions. Accordingly, local and edge subtasks execute in parallel: offloaded ones upload source images for edge preprocessing with outputs directly forwarded to main task, while local ones extract conditions on-device and transmit results to the edge. Once all subtasks complete, the \textit{Conditioning Scale Estimator} estimates the conditioning scale for each visual condition and filters out insignificant ones. The remaining conditions and estimated scales are fed into the main model to generate images, which are then send back to users.


\subsection{Subtask Manager}\label{secsys2}

For each multi-condition T2I generation request from user $k$, it comprises $N_k$ preprocessing subtasks, denoted as $\mathbf{T}_k = \{T_k^1,T_k^2,...,T_k^{N_k}\}$, where $T_k^i$ ($1\leq i \leq N_k$) represents the $i$-th subtask, along with a main image generation task denoted as $T_k^0$. Let $x_k^i \in \{0,1\}$ be a binary variable indicating the offloading decision of subtask $T_k^i$, where a value of 0 denotes local execution on the device, while a value of 1 denotes offloading to the edge for execution.

Given the heterogeneous computational requirements of subtasks, we let $C_k^i$ denote the workload of subtask $T_k^i$ (in FLOPs) and $f_k$ the local processing capability (in FLOPS) of user $k$. The resulting local computing delay for all subtasks of user $k$ is then expressed as
\begin{equation}
	\widehat{L}_k^{\text{U}} = \sum_{i=1}^{N_k} \frac{(1 - x_k^i) C_k^i}{f_k}, \quad \forall k \in \mathcal{K},
\end{equation}
where $\mathcal{K}$ denotes the total user set. Let $f_\text{E}$ denote the computational capacity of the edge server. Accordingly, the edge computing delay for all subtasks of user $k$ can be modeled as
\begin{equation}
	\widehat{L}_{k}^{\text{E}} = \sum_{i=1}^{N_k} \frac{x_k^i C_k^i}{f_\text{E}}, \quad \forall k \in \mathcal{K}.
\end{equation}

Subtasks may run locally or at the edge, creating user-specific uplink demands. To accommodate this heterogeneity, we adopt an orthogonal frequency division multiple access (OFDMA) communication model, where the large number of available subcarriers allows bandwidth to be treated as a continuous variable \cite{bwallo,kongtcsvt}. Let $y_k \in (0,1)$ denote the uplink bandwidth fraction for user $k$. Assuming a frequency-flat channel with channel gain $h_k$ and maximum transmit power $p_k$, the uplink transmission rate of user $k$ is given by the Shannon capacity formula:
\begin{equation}
	r_k = y_k B^{\uparrow} \log \left(1 + \frac{p_k h_k}{y_k B^{\uparrow} N_0}\right), \quad \forall k \in \mathcal{K},
\end{equation}
where $N_0$ is the power of Additive White Gaussian Noise (AWGN), and $B^{\uparrow}$ denotes the total available uplink bandwidth.

Afterwards, we denote by $\widetilde{D}_k^i$ the source image size of subtask $T_k^i$ and $D_k^i$ the size of its generated output. For subtasks offloaded to the edge, the source image must be transmitted from the local for preprocessing, whereas the output generated at the edge is directly forwarded to subsequent generation stages, incurring no additional communication overhead. Thus, the \textit{edge subtask completion time} for user $k$, consisting of transmission and then computation delays, is expressed as
\begin{equation}
	L_k^{\text{E}} =  \sum_{i \in \mathcal{U}_k} \frac{ \widetilde{D}_k^i}{r_k} + \widehat{L}_{k}^{\text{E}}, \quad \forall k \in \mathcal{K},
\end{equation}
where $\mathcal{U}_k$ is the index set of unique input images among the offloaded subtasks of user $k$, such that shared inputs incur the transmission cost only once.

In contrast, when a subtask is executed locally on the user device, the generated output must be transmitted to the edge server after local inference. Therefore, the \textit{local subtask completion time} for user $k$, consisting of computation and then transmission delays, can be expressed as
\begin{equation}
L_k^{\text{U}} =\widehat{L}_{k}^{\text{U}} +  \sum_{i=1}^{N_k} \frac{(1 - x_k^i) D_k^i}{r_k}, \quad \forall k \in \mathcal{K}.
\end{equation}


Since local and edge computing resources operate independently, user-side and edge-side subtasks can be executed in parallel. The main task, however, can only be initiated once all output conditions have arrived at the edge. Accordingly, the subtask completion time for user $k$ is governed by the slower side and is defined as $L_k=\max(L_k^{\text{U}}, L_k^{\text{E}})$. 

\subsubsection{Problem Formulation}
Our objective is to minimize  $L_k$ for each user through joint offloading decisions and bandwidth allocation, thereby accelerating the response of the main generation task. Therefore, the problem is formulated as
\begin{align}
	\mathbf{P}_0: &\quad \min\limits_{\{\mathbf{x,y}\}} \frac{1}{K} \sum_{k \in \mathcal{K}} L_k \\
	\text{s.t.} \quad &x_k^i \in \{0,1\}, \quad \forall k \in \mathcal{K}, 1 \leq i \leq N_k, \label{c1}\\
	&y_k \in (0,1), \quad \sum_{k \in \mathcal{K}}y_k = 1, \quad \forall k \in \mathcal{K}, \label{c2}\\
	&\sum_{i \in N_k} (1 - x_k^i) C_k^i \le \tilde{C}_k, \quad \forall k \in \mathcal{K},\label{c3}\\
	&C_0 + \sum_{k \in \mathcal{K}} \sum_{i \in N_k} x_k^i C_k^i \le \tilde{C}_\text{E},\label{c4}
\end{align}
where $\tilde{C}_k$ and $\tilde{C}_\text{E}$ denotes the available computing resources of user $k$ and at the edge server, respectively, and $C_0$ denotes the computing resources reserved for the main generation task to serve all users. Due to limited device capability, the total workload of locally executed subtasks must not exceed $\tilde{C}_k$. Similarly, the edge server is also subject to capacity constraints, as it must concurrently execute conditional subtasks and the main generation model for multiple users.

The problem $\mathbf{P}_0$ is a Mixed-Integer Nonlinear Programming (MINLP) problem, whose solution space grows exponentially with the number of subtasks, leading to an exponential-time complexity of $\mathcal{O}(2^{\sum_k N_k})$ for brute-force search. Thus, we develop an efficient algorithm to tackle this problem. 


\begin{algorithm}[t] 
  \small
	\caption{Subtask Management Algorithm}
	\KwIn{ $K, f_{\text{E}}, \tilde{C}_{\text{E}}, B^{\uparrow}, \{f_k, \tilde{C}_k, N_k\}_{k \in \mathcal{K}}, L_{0}.$\\
            }
	\KwOut{ Optimized solutions
		$\mathbf{x}^*$, $\mathbf{y}^*$.
            }

	Initialize $\mathbf{y} \leftarrow (1/K, \dots, 1/K)$, $\mathbf{x} \leftarrow \mathbf{0}$

	\For{$iter=1$ \textbf{to} MAX\_ITER}{
		\For{$k \in \mathcal{K}$}{
			Compute user/edge execution latency $L_{k}^{\text{U}}, L_{k}^{\text{E}}$ 
			

			\While{$L_{k}^{\text{U}}-L_{k}^{\text{E}}>\frac{L_k}{N_k}$ \textbf{or} \text{Constraint} \eqref{c3}}{

				$i^* \leftarrow \underset{i: x_{k}^i = 0}{\arg\max} \{\Delta L_k = L_k|_{x_{k}^i = 0} - L_k|_{x_{k}^i = 1} \}$
				
				$x_{k}^{i^{*}} \leftarrow 1$ \textbf{if} $\Delta L_k > 0$ \textbf{and} \textit{Constraint} \eqref{c4}
			}

			$y_k \leftarrow  \underset{y \in (0,1)}{\arg\min}\{L_k \leq L_0\}$

			$\omega_k \leftarrow  \max\{(L_{k}^{\text{E}}-L_{k}^{\text{U}}),0\} * L_k$
		}

		\If{$\sum_{k \in \mathcal{K}}y_k \le 1$}{ 
        
		$\mathbf{y^*} \leftarrow \mathbf{y} + (1-\sum_{k \in \mathcal{K}}y_k) \frac{\bm{\omega}}{\sum_{k \in \mathcal{K}}{\omega_k}} $

        $L_{0} \leftarrow L_{0} - \underset{k \in \mathcal{K}}{\min} \frac{L_k}{N_k}$

		\If{$F(\mathbf{x^*}, \mathbf{y^*}) > F_{prev}(\mathbf{x}, \mathbf{y})$}{
			Accept $\mathbf{x^*}, \mathbf{y^*}$, \quad $F_{prev}(\mathbf{x}, \mathbf{y}) \leftarrow F(\mathbf{x^*}, \mathbf{y^*})$
        
		}
		}

		\Else{
		
		$L_{0} \leftarrow L_{0} + \underset{k \in \mathcal{K}}{\min} \frac{L_k}{N_k}$
		}

    } 

	\Return $\mathbf{x^*}, \mathbf{y^*}$

\label{algo}
\end{algorithm}

\subsubsection{Algorithm Design}

As summarized in Algorithm 1, our algorithm initializes with uniform uplink bandwidth allocation and executes all subtasks locally. In each iteration, the algorithm evaluates the local and edge execution delays $L_{k}^{\text{U}}, L_{k}^{\text{E}}$ for each user and determines the current subtask completion time $L_k$. If local execution dominates the overall latency by exceeding the edge delay beyond the average per-subtask latency, or if the user's available computing resources is inadequate, the algorithm identifies and offloads the subtask that offers the maximum marginal latency reduction, subject to positive latency gain and feasible edge resource constraint.

After updating the offloading configuration, the algorithm computes for each user the minimum uplink bandwidth $y_k$ required to meet the latency constraint $L_0$. We assign a weight $\omega_k$ to quantify the severity of each user's communication bottleneck. Specifically, $\omega_k$ increases when $L_{k}^{\text{E}} > L_{k}^{\text{U}}$, indicating communication possibly dominates the latency, and is further scaled by $L_k$ to prioritize users with higher subtask completion delays. Once the minimal requirements are met, any remaining bandwidth is redistributed proportionally to $\omega_k$, with $L_0$ tightened to the minimal average subtask completion time to enhance efficiency. If feasibility cannot be maintained, $L_0$ is relaxed similarly. The updated result $ (\mathbf{x^*}, \mathbf{y^*})$ is accepted only if it improves the global objective.




\subsection{Conditioning Scale Estimator}\label{secsys4}

After all subtasks complete, multiple visual conditions are produced for the main generation task. As discussed in Section \ref{secmoti2}, appropriate conditioning scales are essential since default scales can over-constrain the model and degrade quality, while removing insignificant conditions can improve efficiency.


To this end, we propose a feature-driven \textit{Conditioning Scale Estimator} that automatically assigns appropriate conditioning scales. During inference, we first load only the lightweight conditioning encoder of each ControlNet branch, which is decoupled from the computationally expensive control blocks and zero-convolution layers. The extracted high-level features are then used to estimate the conditioning scale and determine whether to activate the full ControlNet branch. 

Specifically, we consider a generation with $J$ visual conditions. The $j$-th condition image is first encoded into a high-dimensional feature via its corresponding ControlNet conditioning encoder, then spatially downsampled and channel-wise normalized to obtain the final feature representation $E_j$.

Visual conditions exhibit spatially varying influence as some regions effectively guide generation, while others provide little information. Therefore, we introduce the \textit{spatial intensity map} for the $j$-th visual condition as $A_j=norm(\frac{1}{P}\sum_{p}(E_{j,p})^2)$, where $p$ indexes the channels. The resulting map captures the intensity across spatial regions. However, feature activation intensity alone is insufficient. It is equally important to evaluate whether the activated regions exhibit meaningful spatial variations. Accordingly, we define the \textit{effectiveness score} $e_j$ for the $j$-th visual condition as the average intensity-weighted spatial variance across all channels:
\begin{equation}
  e_j = \frac{1}{P}\sum_{p}\sum_{h,w} A_j(E_{j,p}(h,w)-\overline{E_{j,p}})^2, \quad \forall j \in \mathcal{J}.
\end{equation}
This metric highlights meaningful structural variations in high-intensity regions while suppressing background or noisy fluctuations. A higher $e_j$ indicates that the corresponding visual condition exhibits not only strong activations but also pronounced spatial variations, enabling it to exert more effective constraints on the latent representation during denoising. 

Moreover, in multi-condition scenarios, some control signals may be highly correlated at the semantic level. Assigning large scales to all conditions indiscriminately can introduce redundancy. To alleviate this issue, we propose a soft redundancy suppression mechanism by defining the \textit{uniqueness score} as
\begin{equation}
  u_j = -\frac{1}{J-1}\sum_{m \neq j} \max(0,\cos(\mathbf{v}_j, \mathbf{v}_m)-\delta), \quad \forall j \in \mathcal{J},
\end{equation}
where $\mathbf{v}_j=norm(\sum_{h,w}A_j E_j)$ is the feature vector. Only when the cosine similarity exceeds a threshold $\delta$, a penalty term is imposed to suppress redundant conditions, encouraging complementarity while avoiding excessive competition.

Finally, we jointly model the condition effectiveness and uniqueness to estimate the conditioning scale $\alpha_j$ through a weight parameter $\lambda$. Conditions with low scores are regarded as insignificant, which are pruned via by setting $\alpha_j=0$ when it falls below a threshold $\theta$. To summarize, the estimated conditioning scale is defined using an indicator function as
\begin{equation}
  \alpha_j = (\lambda u_j + (1-\lambda) e_j) \cdot \mathbb{I}((\lambda u_j + (1-\lambda) e_j) \ge \theta).
\end{equation}
By pruning insignificant conditions, the corresponding ControlNet branches are neither loaded nor executed, effectively reducing the denoising latency while preserving quality.

\section{Performance Evaluation}\label{seceval}


\subsection{Experimental Settings}\label{secval1}

\subsubsection{Implementation} We consider one edge server equipped with an NVIDIA GeForce RTX 3080Ti GPU serves five users. The five users are emulated using one Jetson AGX Orin, two Jetson TX2 and two Jetson Nano devices. Stable Diffusion v1.5 and ControlNet models \cite{controlnet} are utilized for generation with 20 denoising steps. We properly normalize DreamSim \cite{dreamsim} and ImageReward \cite{imagereward} metrics for quality assessment. We set $B^{\uparrow}$=20 MHz, $L_0$=1s, $\lambda$=0.2, $\delta$=0.6, $\theta$=0.2 by default. We conduct experiments on a subset of \textit{Subjects200K} dataset \cite{subject200k} and the number of subtasks per user ranges from 1 to 5. 


\subsubsection{Benchmarks} We compare with these baselines.

\begin{itemize}[]
	\item {\textbf{\textit{Local}}}. All subtasks are executed locally, except when limited local resources necessitate edge execution.
	\item {\textit{\textbf{Edge.}}} \cite{swtmc}. All subtasks are executed at edge, except when limited edge resources necessitate local execution.
  \item {\textit{\textbf{Random.}}} Offloading decisions, bandwidth allocations and conditioning scales are selected randomly.
	\item {\textbf{\textit{ComDOS}}} \cite{comdos}. The entire generation process is modeled as a DAG, and dynamic programming is applied to select the offloading strategy that maximizes system utility. 
	\item {\textbf{\textit{AMARL}}} \cite{amarl}. An attention-enhanced multi-agent reinforcement learning approach is used to optimize offloading decisions that minimize overall system latency.
\end{itemize}

\subsection{Overall Performance}\label{secval2}

\subsubsection{Latency analysis}
We first evaluate the overall latency of different methods in terms of average subtask computation, subtask transmission, and main task inference, as shown in Fig. \ref{figrelatency}. The \textit{Local} scheme has minimal transmission latency but suffers from prolonged computation due to limited device capabilities, while the \textit{Edge} scheme minimizes computation latency at the cost of high transmission overhead from uploading raw images. \textit{ComDOS} reduces computation latency via dynamic programming but ignores multi-user bandwidth contention, limiting transmission optimization. \textit{AMARL}, while learning offloading policies, overlooks heterogeneous transmission demands and may trigger resource overload, resulting in 2.55 s latency. Our method jointly optimizes subtask offloading and bandwidth allocation, enabling parallel end-edge execution to complete in 0.57 s, and further reduces overall latency to 2.14 s by pruning insignificant visual conditions.


\subsubsection{Quality analysis}

\begin{figure}[t]
   \centering
   \begin{minipage}[t]{0.24\textwidth}
     \centering
     \includegraphics[width=\textwidth]{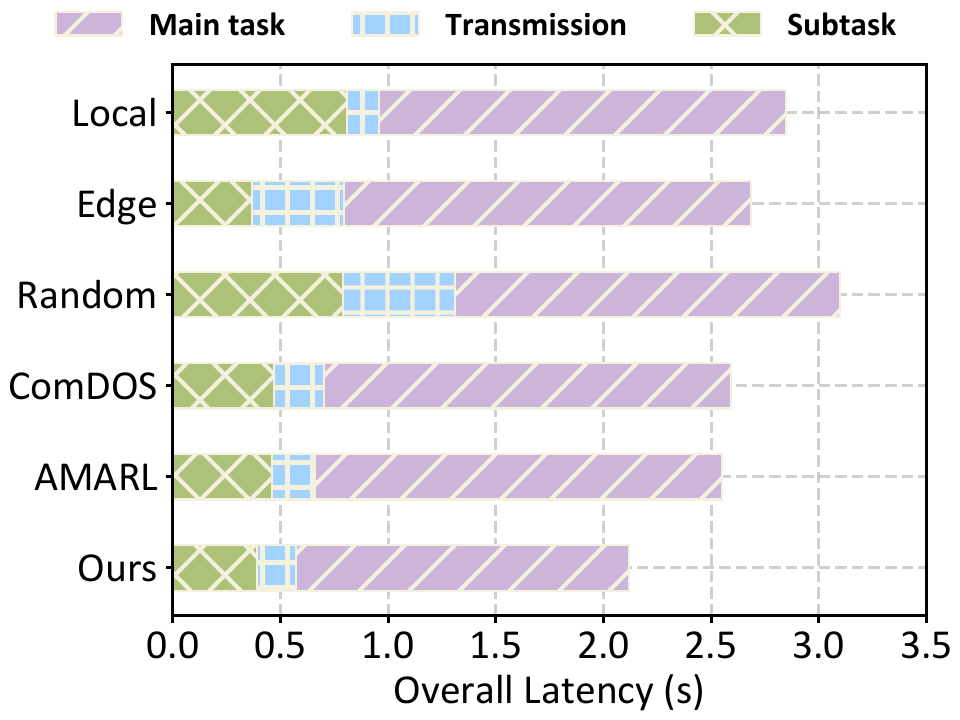}
     \caption{Overall latency analysis.}
     \label{figrelatency}
   \end{minipage}
   \hfill
   \begin{minipage}[t]{0.24\textwidth}
     \centering
     \includegraphics[width=\textwidth]{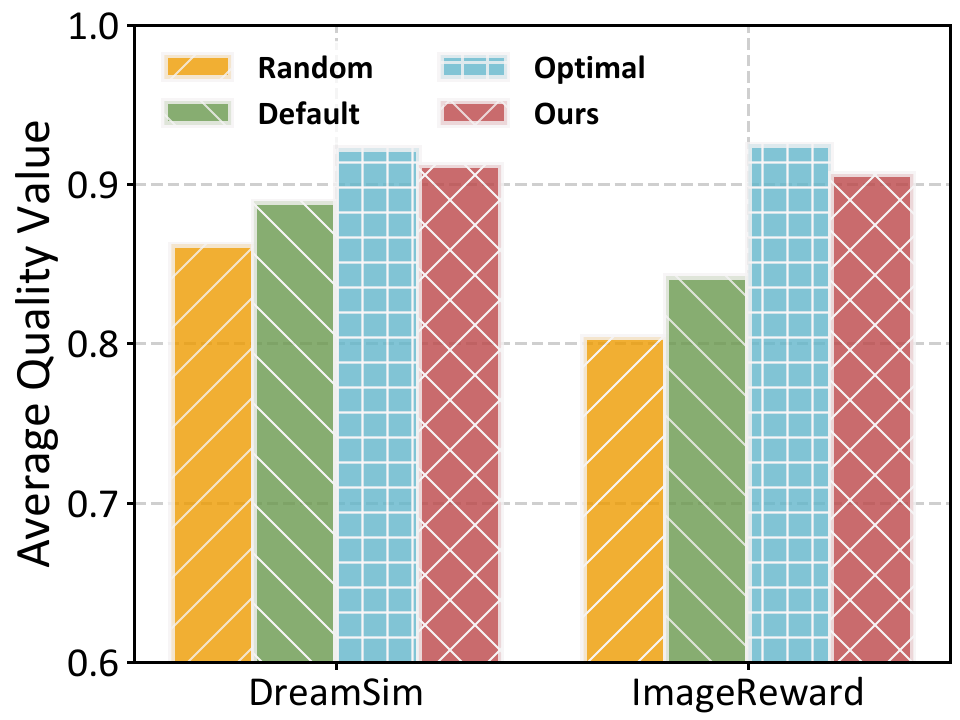}
     \caption{Generation quality analysis.}
     \label{figrequality}
   \end{minipage}
 \end{figure}

\begin{figure}[t]
  \centering
  \includegraphics[scale=0.46]{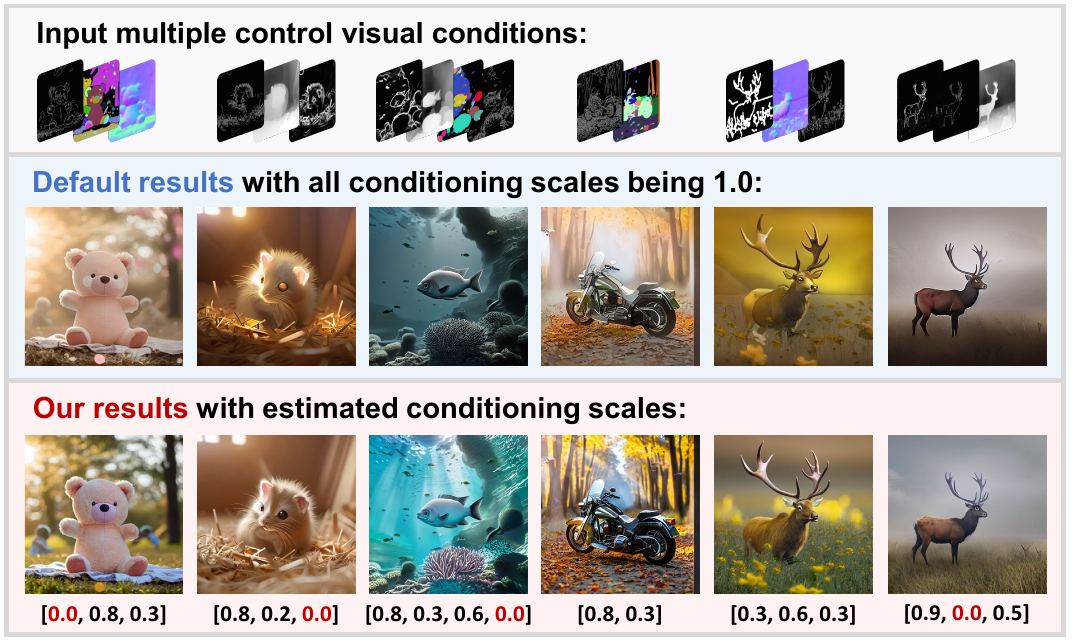}
  \caption{Visualization comparison under different conditioning scales.}
  \label{figrevis}
\end{figure}

Next, we evaluate the generation quality under different schemes, as depicted in Fig. \ref{figrequality}. We compare the estimated conditioning scales from our method against the \textit{default} scheme (\textit{i.e.,} all scales set to 1), the \textit{random} scheme, and the \textit{optimal} scheme (\textit{i.e.,} exhaustively searched for best quality). Compared to the default scheme, our method adaptively sets conditioning scales according to inter-condition similarity and informativeness, effectively filtering out low-contribution conditions, thus achieving 2\% and 6\% gains in DreamSim and ImageReward respectively. Relative to the optimal configuration, our method incurs only a 1\% drop in DreamSim and a 2\% drop in ImageReward, demonstrating its effectiveness. Moreover, several visual comparisons between our method and the default approach are presented in Fig. \ref{figrevis}.

 \begin{figure}[t]
  \centering
  \begin{minipage}[t]{0.24\textwidth}
    \centering
    \includegraphics[width=\textwidth]{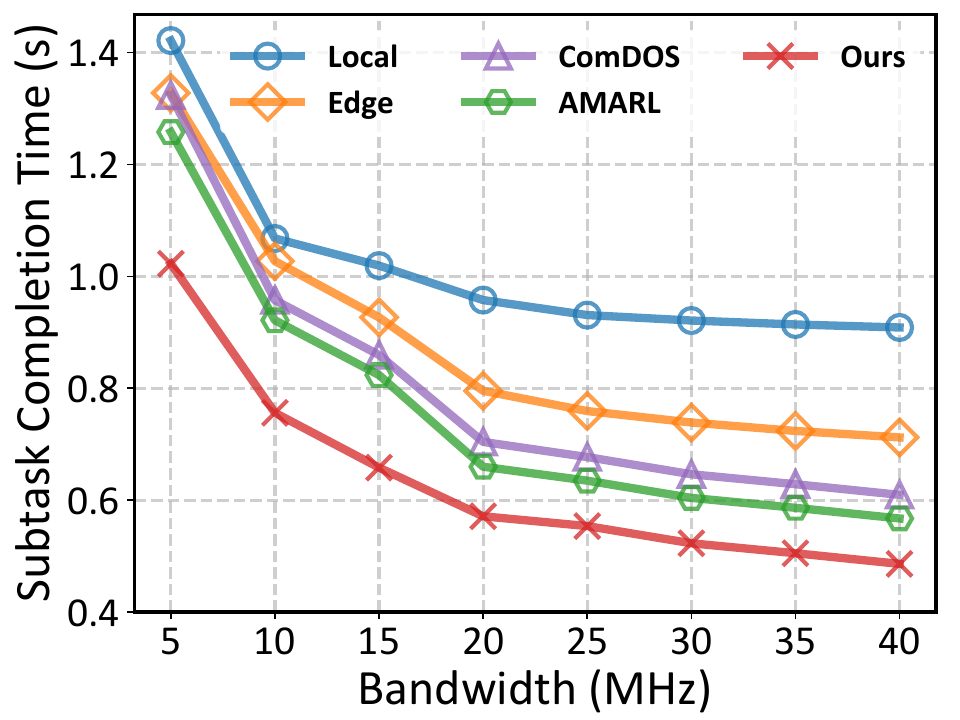}
    \caption{Impact of bandwidth $B^{\uparrow}$.}
    \label{figrebw}
  \end{minipage}
  \hfill
  \begin{minipage}[t]{0.24\textwidth}
    \centering
    \includegraphics[width=\textwidth]{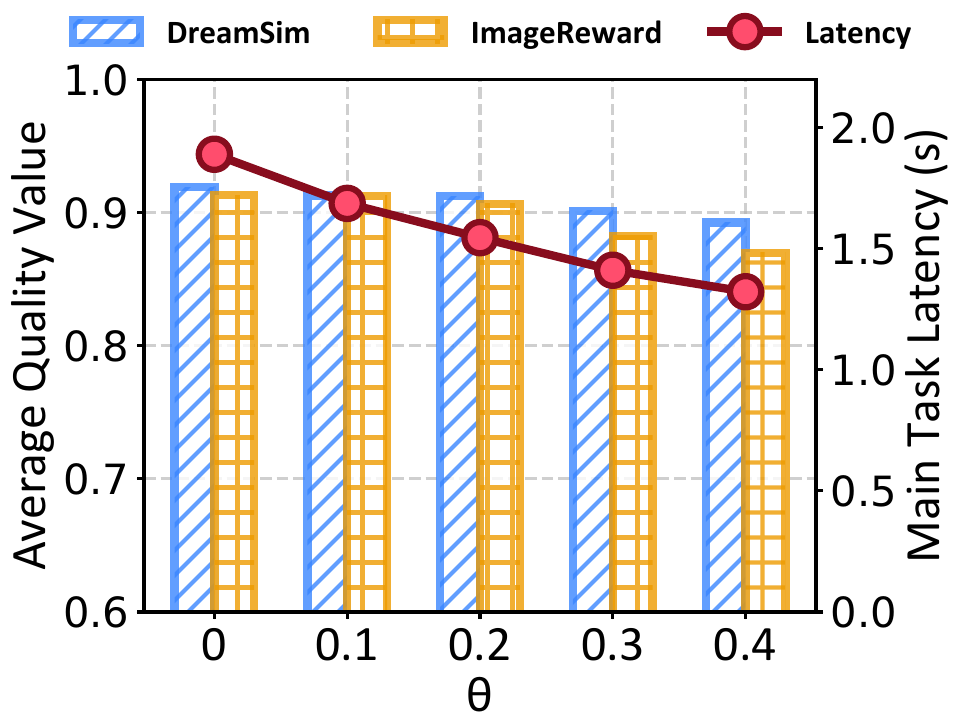}
    \caption{Impact of threshold $\theta$.}
    \label{figrethre}
  \end{minipage}
\end{figure}

\subsection{Impact of Parameters}\label{secval3}

\subsubsection{Impact of bandwidth resource} 
As illustrated in Fig. \ref{figrebw}, subtask completion latency decreases with increasing uplink bandwidth for all schemes, primarily due to alleviated multi-user contention. Our approach consistently yields the lowest latency because of our consideration of heterogeneous user transmission demands in the subtask management algorithm, which enables joint optimization of the bandwidth allocation with the offloading strategy even under severe bandwidth constraints. For instance, when the available bandwidth is only 5 MHz, our method still reduces latency by about 0.4 s.


\subsubsection{Impact of filtering threshold}
We then assess the effect of the filtering threshold $\theta$ on main task inference latency and image quality, as shown in Fig. \ref{figrethre}. When $\theta = 0$ (no filtering), latency matches the baseline at 1.89 s. Decreasing $\theta$ eliminates more visual conditions, thereby reducing the number of active ControlNet branches and shortening inference time. However, this acceleration compromises quality. At $\theta = 0.4$, ImageReward score decreases by about 5\% as critical conditions are inadvertently pruned. We adopt $\theta = 0.2$ to strike a favorable balance between latency reduction and quality preservation.


\section{Conclusion}\label{seccon}

In this paper, we have proposed an end-edge collaborative system design to accelerate multi-condition T2I generation. By jointly optimizing subtask offloading and bandwidth allocation through the \textit{Subtask Manager}, and pruning insignificant visual conditions via the \textit{Conditioning Scale Estimator}, our system reduces average latency by 25\% while improving generation quality by 6\%. The outcome of this paper can help to pave the way for more interactive and controllable generative applications at the edge. Future work will extend this system design to multi-edge collaboration to support larger-scale deployments.



\bibliographystyle{IEEEtran}
\bibliography{ref}

\end{document}